# Detection of picosecond magnetization dynamics of 50 nm magnetic dots down to the single dot regime

Bivas Rana[1], Dheeraj Kumar[1], Saswati Barman[1], Semanti Pal[1], Yasuhiro Fukuma[2], YoshiChika Otani[2,3*] and Anjan Barman[1,†]

[1]*Department of Condensed Matter Physics and Material Sciences, S. N. Bose National Centre for Basic Sciences, Block JD, Sector III, Salt Lake, Kolkata 700 098*

[2]*Advanced Science Institute, RIKEN, 2-1 Hirosawa, Wako, Saitama 351-0198, Japan*

[3]*Institute for Solid State Physics, University of Tokyo, 5-1-5 Kashiwanoha, Kashiwa, Chiba 277-8581, Japan*

*CORRESPONDING AUTHOR, EMAIL ADDRESS: yotani@issp.u-tokyo.ac.jp

[†]abarman@bose.res.in

**ABSTRACT:** We report an all-optical time-domain detection of picosecond magnetization dynamics of arrays of 50 nm $Ni_{80}Fe_{20}$ (permalloy) dots down to the single nanodot regime. In the single nanodot regime the dynamics reveals one dominant resonant mode corresponding to the edge mode of the 50 nm dot with slightly higher damping than that of the unpatterned thin film. With the increase in areal density of the array both the precession frequency and damping increases significantly due to the increase in magnetostatic interactions between the nanodots and a mode splitting and sudden jump in apparent damping are observed at an edge-to-edge separation of 50 nm.

KEYWORDS: Single Nanomagnets Dynamics, Time-resolved Magneto-optical Kerr Effect, Magnetization Precession, Damping, Magnetostatic Interaction.

The quest to measure the ultrafast magnetization dynamics of nanomagnets continues to be an important problem in nanoscience and nanotechnology.[1-9] Picosecond magnetization dynamics of nanoscale magnetic structures is important for many present and future technologies including magnetic data storage,[10-11], logic devices,[12-14] spintronics,[15] and magnetic resonance imaging.[16] Emerging technologies such as spin torque nano-oscillators[17] and magnonic crystals[18-19] rely heavily upon the fast and coherent dynamics of nanomagnets and the generation and manipulation of spin waves in spatially modulated magnetic nanostructures. Novel techniques for fabrication of nanomagnets arrays[20] and applications towards biomedicine[21] show exciting new promises. Overall, the detection and understanding of nanomagnet dynamics down to the single nanomagnet regime have become increasingly important. Investigation of picosecond dynamics of arrays of nano-scale magnetic dots has inferred that, for dot sizes less than 200 nm, the response of the magnetization to a pulsed magnetic field is spatially non-uniform and is dominated by localized spin wave modes.[22] This non-uniformity may result in a degradation of the signal to noise ratio in future nanomagnetic devices. However, the measurements were done in densely packed arrays where the intrinsic dynamics of the individual dots are strongly influenced by the magnetostatic stray fields of the neighbouring dots. Magnetostatically coupled nanomagnets in a dense array may show collective behaviors both in the quasistatic magnetization reversal[23] and in the precessional dynamics.[22, 24-28] In the quasistatic regime the strong inter-dot magnetostatic interactions result in collective rotation of magnetic spins and formation of flux closure through a number of dots during the reversal. On the other hand, in the collective precessional dynamics the constituent nanomagnets maintain definite amplitude and phase relationships. Magnetization dynamics in dense arrays of nanomagnets have been studied both experimentally by time-domain,[22,24] frequency-domain[25-26] and wave-vector-domain[27-28] techniques; and theoretically by analytical[29-30] and micromagnetic[31] methods. To this end the frequency, damping and spatial patterns of spin waves and dispersion relations of frequency with wave-vector of spin wave propagation have been studied.

On the other hand, magnetization dynamics of isolated nanomagnets with lateral dimensions down to 125 nm have been reported by time-resolved magneto-optical techniques.[3,5,6,8,9] However, picosecond magnetization dynamics including the damping behavior of isolated nanomagnets down to 50 nm size has never been reported. Here, we present an all-optical far field measurement of the picosecond magnetization dynamics of arrays of square $Ni_{80}Fe_{20}$ (permalloy) dots with 50 nm width and with varying edge-to-edge separation ($S$) between 200 nm and 50 nm. When the dots are separated by large distance ($S \geq 150$ nm) they reveal the dynamics of the isolated nanomagnet. The isolated nanomagnets revealed a single resonant mode, whose damping is slightly higher than the unpatterned thin film value. With the decrease in inter-dot separation the effects of dipolar and quadrupolar interactions become important, and we observe an increase in precession frequency and damping. At the highest areal density a sudden jump in the apparent damping is observed due to the mutual dephasing of two closely spaced eigenmodes of the array.

**RESULTS AND DISCUSSION**

10 μm × 10 μm square arrays of permalloy dots with nominal dimensions as 50 nm width, 20 nm thickness and separation $S$ varying from 50 nm to 200 nm were prepared by a combination of electron beam evaporation and electron-beam lithography. Figure 1(a) presents the scanning electron micrographs of three of these dot arrays, which show that there are some deviations in the shape and dimensions of the samples from the nominal shape and dimensions, although the general features are maintained. A square permalloy dot with 10 μm width and 20 nm thickness was also prepared to obtain the magnetic parameters of the unpatterned sample. The ultrafast magnetization dynamics was measured by using a home-built time-resolved magneto-optical Kerr effect microscope based upon a two-color collinear pump-probe set-up.[32] The two-color collinear arrangement enabled us to achieve a very good spatial resolution and sensitivity even in an all-optical excitation and detection scheme of the precessional dynamics. A schematic of the measurement geometry is shown in Fig. 1(b). The time-resolved data was recorded for a maximum duration of 1 ns and this was found to be sufficient to record all important features of the dynamics including the spectral resolution of the double peaks for the sample with $S$ = 50 nm and measurement of the damping coefficient. Figure 1(c) shows the time-resolved reflectivity and Kerr rotation data from the array with separation $S$ = 50 nm at a bias field $H$ = 2.5 kOe. The reflectivity shows a sharp rise followed by a bi-exponential decay. On the other hand the time-resolved Kerr rotation shows a fast demagnetization within 500 fs and a bi-exponential decay with decay constants of about 8 ps and 116 ps. The demagnetization and decay times are found to be independent of the areal density of the arrays. The precessional dynamics appears as an oscillatory signal[2] above the decaying part of the time-resolved Kerr rotation data. The bi-exponential background is subtracted from the time-resolved Kerr signal before performing the fast Fourier transform (FFT) to find out the corresponding power spectra.

Figure 2 shows the time-resolved Kerr rotation from the permalloy dot arrays with $S$ varying between 50 nm and 200 nm at $H$ = 2.5 kOe. Clear precession is observed down to $S$ = 200 nm, where the dots are expected to be magnetostatically isolated and hence exhibit single dot like behavior. The corresponding FFT spectrum (Fig. 2(b)) shows a dominant single peak at 9.04 GHz. As $S$ decreases the precession continues to have a single resonant mode but the peak frequency generally increases with the decrease in $S$. For $S$ = 150 nm, the peak frequency decreases slightly although the errors bars are large enough to maintain the general trend of increase in the frequency with the decrease in $S$ as stated above. At $S$ = 50 nm the single resonant mode splits into two closely spaced modes with the appearance of a lower frequency peak. In Fig. 2(c), we show the FFT spectra of the time-resolved magnetization obtained from micromagnetic simulations of arrays of 7 × 7 dots using the OOMMF software.[33] In general, the deviation in the shape and dimensions as observed in the experimental samples are included in the simulated samples but the precise edge roughness profiles and deformations are not always possible to include in the finite difference method based micromagnetic simulations used here, where samples are divided into rectangular prism like cells. In the simulation the arrays were divided into cells of 2.5 × 2.5 × 20 nm³ dimensions and material's parameters for permalloy were used as $\gamma = 18.5$ MHz/Oe, $H_K = 0$, $M_S = 860$ emu/cc and $A = 1.3 \times 10^{-6}$ erg/cm. The material's parameters for permalloy were obtained by measuring the precession frequency of the unpatterned thin film as a function of the in-plane bias field and by fitting the bias field variation of frequency with Kittel's formula. The exchange stiffness constant $A$ was obtained from literature.[34] The lateral cell size is well below the exchange length $l_{ex} = \sqrt{\frac{2A}{\mu_0 M_S^2}}$ of permalloy (5.3 nm) and further reduction of cell size does not change the magnetic energies appreciably. Test simulations with discretization along the thickness of the samples do not show any variation in the resonant modes, which is expected as this will only affect the perpendicular standing spin waves, whereas in the present study we have concentrated on the spin-waves with in-plane component of wave-vector. The equilibrium states are obtained by allowing the system to relax under the bias field for sufficient time so that the maximum torque ($m \times H$, $m = M/M_S$) goes well below $10^{-6}$ A/m. The dynamic simulations were obtained for a total duration of 4 ns at time steps of 5 ps. Consequently, the simulated linewidths of the resonant modes are narrower, which enabled us to clearly resolve the mode splitting in the simulation. The simulation reproduces the important features as observed in the experiment, namely the observation of a single resonant mode for the arrays with $S$ varying between 200 nm and 75 nm, a systematic increase in the resonant mode frequency with the decrease in $S$, and finally a mode splitting at $S = 50$ nm. However, the increase in the resonant frequency with decrease in $S$ is less steep as compared to the experimental result. The deviation is larger for smaller values of $S$ possibly due to the increased non-idealities in the physical structures of the samples in this range, as discussed earlier. Furthermore the relative intensities of the two modes observed for the array with $S = 50$ nm are not reproduced by the simulation. This is possibly because the lower frequency mode is a propagating mode and the finite boundary of the simulated array of 7 × 7 elements may cause much faster decay of the propagating mode as opposed to that in the much larger array of 100 × 100 elements studied experimentally. In Fig. 3(a), we plot the precession frequency as a function of the ratio of width ($w$) to centre to centre separation ($a$), where $a = w + S$. For $w/a \leq 0.25$ ($S \geq$ 150 nm) the frequency is almost constant but for $w/a > 0.25$ ($S < 150$ nm) the frequency increases sharply both for the experimental and simulated data. We fit both data with Eq. 1 including both dipolar and quadrupolar interaction terms.[35]

$$f = f_0 - A\left(\frac{w}{a}\right)^3 + B\left(\frac{w}{a}\right)^5 \qquad [1],$$

where $A$ and $B$ are the strengths of the dipolar and quadrupolar interactions. The fitted data are shown by solid lines in Fig. 3(a). The simulated data fits well with Eq. 1, while the fit is reasonable for the experimental data, primarily due to the large deviation in data points for the arrays with $S = 75$ and 150 nm. However, the theoretical curve passes through the error bars for those data points. The quadrupolar contribution is dominant over the dipolar contribution as is also evident from the sharp increase in the frequency for $w/a > 0.25$. The dipolar contributions extracted from the curve fitting are almost identical for both experimental and simulated results, whereas for the experimental data the quadrupolar contribution is about 30% greater than that for the simulated data..

**VARIATION OF DAMPING OF PRECESSION WITH AREAL DENSITY OF THE ARRAYS**

We have further investigated the damping behavior of the nanomagnets in the array. The time domain data was fitted with a damped sine curve

$$M(t) = M(0)e^{\frac{-t}{\tau}} \sin(2\pi ft - \phi) \qquad [2],$$

where the relaxation time $\tau$ is related to the Gilbert damping coefficient $\alpha$ by the relation $\tau = \frac{1}{2\pi f\alpha}$, $f$ is the experimentally obtained precession frequency and $\phi$ is the initial phase of the oscillation. The fitted data is shown by solid lines in Fig. 2(a). The damping coefficient $\alpha$, as extracted from the above fitting, is plotted as a function of the inter-dot separation $S$ along with the error bars in Fig. 3(b). The sample with $S$ = 200 nm shows the lowest $\alpha$ of about 0.023. This value of $\alpha$ is slightly higher than the damping coefficient (0.017) measured for a permalloy film of 20 nm thickness grown under identical conditions to those for the arrays of permalloy dots. Since the dots are magnetostatically isolated, the increase in damping due to the mutual dynamic dephasing of the permalloy dots is unlikely for $S$ = 200 nm. Another possibility is the dephasing of more than one mode within the individual dots,[36] which is also ruled out due to the appearance of a dominant single mode in the individual dots. Hence, we believe this increase in damping is possibly due to the defects[37] produced in these dots during nanofabrication, which is quite likely due to the small size of these dots. As $S$ decreases, the magnetostatic interaction between the dots becomes more prominent and hence the mutual dephasing of slightly out-of-phase magnetization precession of the dots in the array becomes more prominent,[31] and consequently $\alpha$ increases systematically with the decrease in $S$ down to 75 nm. At $S$ = 50 nm a different situation arises, where the single resonant mode splits into two closely spaced modes and the apparent damping (square symbol in Fig. 3(b)) of the time-domain oscillatory signal jumps suddenly from 0.032 to 0.066. Clearly, this is due to the out-of-phase superposition of two closely spaced modes within the array, as shown later in this article. In order to understand the correct damping behavior of the uniform resonant mode we have isolated the time-domain signal for the mode 1 from the lower lying mode (mode 2) by using fast Fourier filtering. The extracted damping of the filtered time-domain signal for the sample with $S$ = 50 nm is about 0.033, which is consistent with the systematic increase in the damping coefficient of the arrays with decreasing $S$, as shown by the circular symbols in Fig. 3(b).

**MICROMAGNETIC ANALYSIS OF THE OBSERVED PRECESSIONAL DYNAMICS**

In order to gain more insight into the dynamics, we have calculated the magnetostatic field distribution of the simulated arrays and the contour plot of the magnetostatic fields from the 3 × 3 dots at the centre of the array is shown in Fig. 4. At larger separations the stray fields from the dots remain confined close to their boundaries and the interactions between the dots is negligible. As the inter-dot separation decreases the stray fields of the neighboring dots start to overlap causing an increase in the effective field acting on the dots and consequently the corresponding precession frequency. At $S = 50$ nm the stray field is large enough to cause a strong magnetostatic coupling between the dots and hence the collective precession modes of the dots in the array.[31] The spatial natures of the modes were investigated by numerically calculating the spatial distributions of amplitudes and phases corresponding to the resonant modes of the samples. The amplitude and phase maps of resonant modes for the arrays with $S = 50$ nm and 200 nm are shown in Figs. 5 (a) - (b). For $S = 50$ nm, the main resonant mode (mode 1) corresponds to the in-phase precession of majority of the dots in the array apart from the dots near the edges. The intensities of the dots increase from the edge to the centre of the array. The lower frequency peak (mode 2), on the other hand, shows that the dots in the consecutive columns precess out-of-phase, while the dots in the alternative columns precess in-phase. The intensity again shows small variation from the edge to the centre of the array. The spatial variation of the phase of precession of the dots is similar to the magnetostatic backward volume modes with the wave-vector parallel to the bias magnetic field ($H$) and both lie within the plane of the sample. For $S = 200$ nm, the single resonant mode (mode 1) corresponds to the precession of the individual dots and hence all of them have identical amplitude and phase. For comparison we have calculated the amplitude and phase maps of the only resonant mode of a single 50 nm wide dot (Figs. 5(c) – (e)) with different cell size ((c): $2.5 \times 2.5 \times 20$ nm$^3$, (d): $1 \times 1 \times 20$ nm$^3$, and (e): $2.5 \times 2.5 \times 5$ nm$^3$), which is found to be the edge mode[22, 35, 38] that occupies the major fraction of the volume of the dot. Important to note that the mode structure remains independent on the chosen cell size. A closer view to the central dot of the array with $S = 200$ nm shows (Fig. 5(f)) an identical mode structure to that of the single dot, ensuring that in this array the dynamics is dominated by that of the single dot.

**CONCLUSIONS**

In summary, we have detected the picosecond precessional dynamics in arrays of 50 nm permalloy dots down to the single nanodot regime by an all-optical time-resolved magneto-optical Kerr effect microscope. The inter-dot separation ($S$) varies from 200 nm down to 50 nm and numerical calculation of magnetostatic fields shows a transition from magetostatically isolated regime to strongly coupled regime as $S$ decreases. Consequently, we observe a single precessional mode for $S$ down to 75 nm, whose frequency increases with the decrease in $S$. This has been analytically modeled by introducing the dipolar and quadrupolar contributions to the precession frequency. At the smallest separation $S = 50$ nm, we observe a splitting of the resonant mode and a lower frequency mode appears in addition to the existing mode. Micromagnetic simulations reproduce the above observations qualitatively. Analyses of amplitude and phase maps of the resonant modes reveal that the dynamics of a single dot with 50 nm width is dominated by the edge mode. In sparsely packed arrays ($S \geq 150$ nm) we primarily observe the isolated dynamics of the constituent dots, all in phase. For $S = 50$ nm, the observed modes correspond to the uniform collective precession of the array (higher frequency mode) and an out-of-phase precession of the alternative columns of the array parallel to the bias field (lower frequency mode). The damping also shows significant variation with the areal density. For $S = 200$ nm *i.e.*, in the single nanodot regime, the damping is minimum at about 0.023, which is slightly higher than the damping coefficient (0.017) of a permalloy thin film of same thickness. We understand this slight increase in damping is a result of the defects introduced in the dots during nanofabrication. However, the damping increases further with the decrease in $S$ as a result of the dynamic dephasing of the precession of the weakly interacting dots. At $S = 50$ nm, the dephasing due to the superposition of two resonant modes results in a sudden increase in the apparent damping of the precession. The ability of all-optical detection of the picosecond dynamics of 50 nm dots down to the single nanomagnet regime and understanding of the effects of magnetostatic interaction on those dots when placed in a dense array will be important from fundamental scientific viewpoint as well as for their future applications in various nanomagnetic devices.

**METHODS**

Square arrays of permalloy dots were prepared by a combination of electron beam evaporation and electron-beam lithography. A bilayer PMMA (poly methyl methacrylate) resist pattern was first prepared on thermally oxidized Si(100) substrate by using electron-beam lithography and permalloy was deposited on the resist pattern by electron-beam evaporation at a base pressure of about $2 \times 10^{-8}$ Torr. A 10 nm thick $SiO_2$ capping layer was deposited on top of permalloy to protect the dots from degradation when exposed to the optical pump-probe experiments in air. This is followed by the lifting off of the sacrificial material and oxygen plasma cleaning of the residual resists that remained even after the lift-off process.

The ultrafast magnetization dynamics was measured by using a home-built time-resolved magneto-optical Kerr effect microscope based upon a two-color collinear pump-probe set-up. The second harmonic ($\lambda$ = 400 nm, pulse width ~ 100 fs) of a Ti-sapphire laser (Tsunami, SpectraPhysics, pulse-width ~ 80 fs) was used to pump the samples, while the time-delayed fundamental ($\lambda$ = 800 nm) laser beam was used to probe the dynamics by measuring the polar Kerr rotation by means of a balanced photo-diode detector, which completely isolates the Kerr rotation and the total reflectivity signals. The pump power used in these measurements is about 8 mW, while the probe power is much weaker and is about 1.5 mW. The probe beam is focused to a spot size of 800 nm and placed at the centre of each array by a microscope objective with numerical aperture N. A. = 0.65 and a closed loop piezoelectric scanning x-y-z stage. The pump beam is spatially overlapped with the probe beam after passing through the same microscope objective in a collinear geometry. Consequently, the pump spot is slightly defocused (spot size ~ 1 μm) on the sample plane, which is also the focal plane of the probe spot. The probe spot is placed at the centre of the pump spot as shown in Fig. 1(b). A large magnetic field is first applied at a small angle (~ 15°) to the sample plane to saturate its magnetization. The magnetic field strength is then reduced to the bias field value ($H$ = component of bias field along x-direction), which ensures that the magnetization remains saturated along the bias field direction. The bias field was tilted 15° out of the plane of the sample to have a finite demagnetizing field along the direction of the pump pulse, which is eventually modified by the pump pulse to induce precessional magnetization dynamics within the dots. The pump beam was chopped at 2 kHz frequency and a phase sensitive detection of the Kerr rotation was used.

*Acknowledgement* The authors gratefully acknowledge the financial supports from Department of Science and Technology, Government of India under the grant numbers SR/NM/NS-09/2007, SR/FTP/PS-71/2007, INT/EC/CMS (24/233552) and INT/JP/JST/P-23/09 and Japan Science and Technology Agency Strategic International Cooperative Program under the grant numbers 09158876.

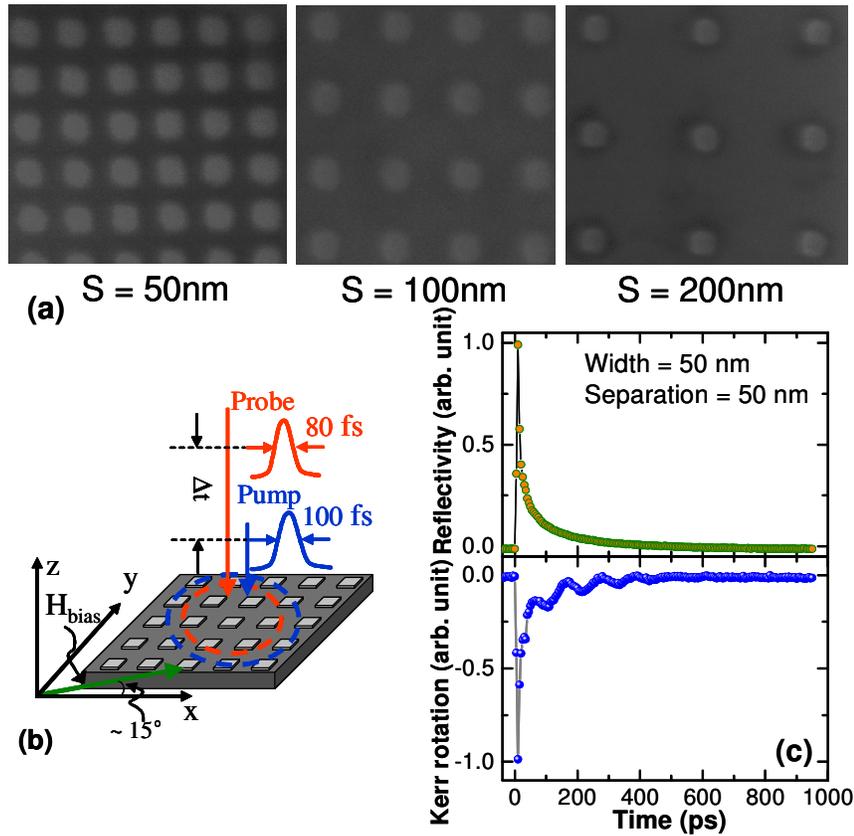

**Figure 1.** (a) Scanning electron micrographs of arrays of permalloy dots with width = 50 nm, thickness = 20 nm and with varying separation $S$ = 50 nm, 100 nm, and 200 nm. (b) A schematic of the two color pump-probe measurement of the time-resolved magnetization dynamics of the nanomagnets. (c) Typical time-resolved reflectivity and Kerr rotation data are shown for the array with $S$ = 50 nm at $H$ = 2.5 kOe.

B. Rana et al.

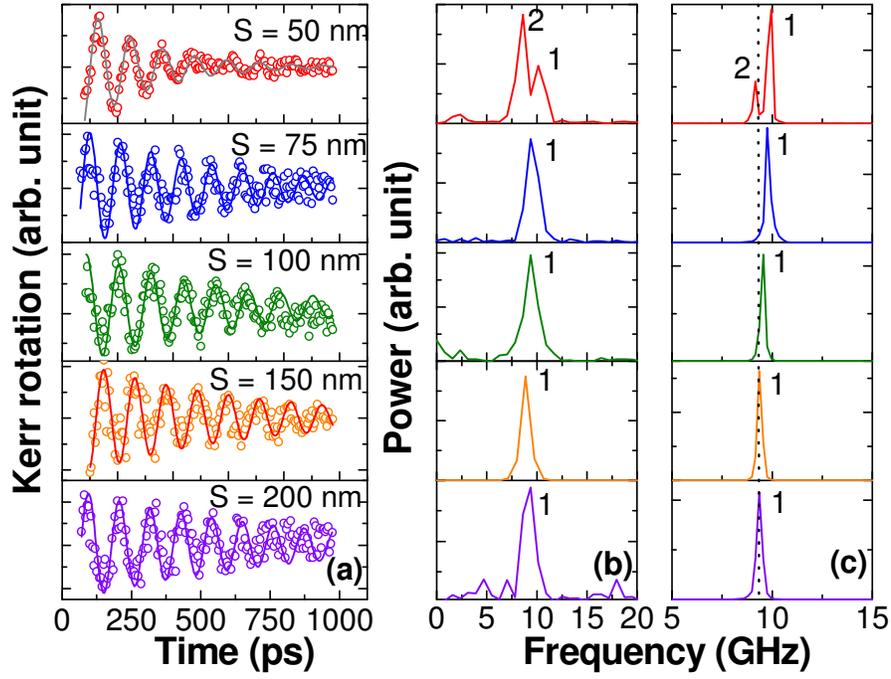

**Figure 2.** (a) Experimental time-resolved Kerr rotations and (b) the corresponding FFT spectra are shown for arrays of permalloy dots with width = 50 nm, thickness = 20 nm and with varying inter-dot separation $S$ at $H$ = 2.5 kOe. (c) The FFT spectra of the simulated time-resolved magnetization are shown. The peak numbers are assigned to the FFT spectra. The dotted line in (c) shows the simulated precession frequency of a single permalloy dot with width = 50 nm, thickness = 20 nm.

B. Rana et al.

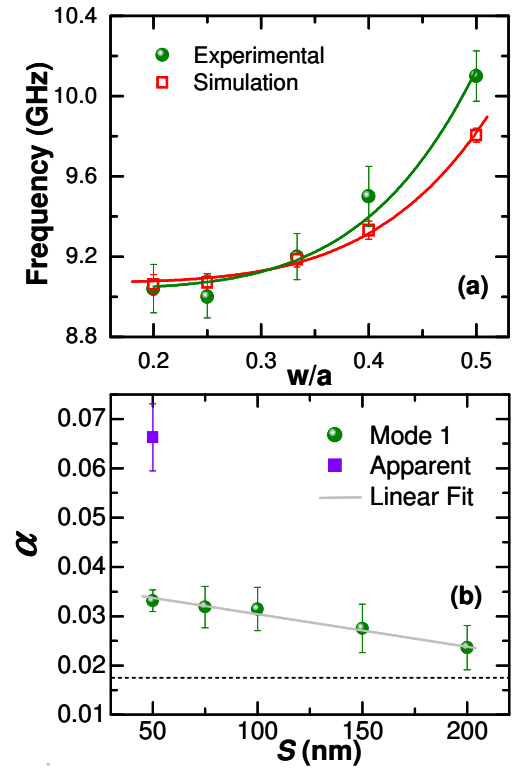

**Figure 3.** (a) The precession frequency is plotted as a function of *w/a*. The circular and square symbols correspond to the experimental and simulated results, respectively, while the solid curves correspond to the fit to Eq. 1. (b) The damping coefficient $\alpha$ is plotted as a function of *S*. The symbols correspond to the experimental data, while the solid line corresponds to a linear fit. The dashed line corresponds to the measured value of $\alpha$ for a continuous permalloy film grown under identical conditions.

B. Rana et al.

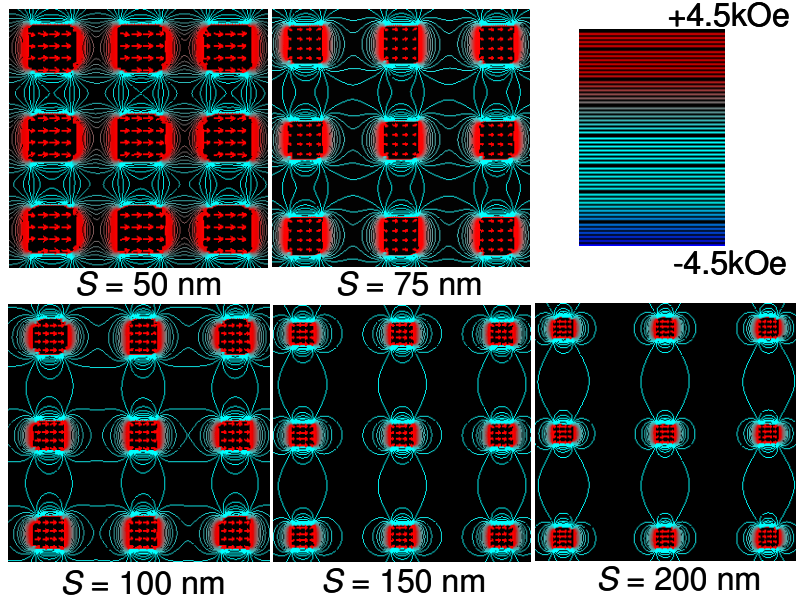

**Figure 4.** Simulated magnetostatic field distributions (x-component) are shown for arrays of permalloy dots with $S$ = 50 nm, 75 nm, 100 nm, 150 nm and 200 nm at $H$ = 2.5 kOe. The arrows inside the dots represent the magnetization states of the dots, while the strengths of the stray magnetic fields are represented by the color bar at the top right corner of the figure.

B. Rana et al.

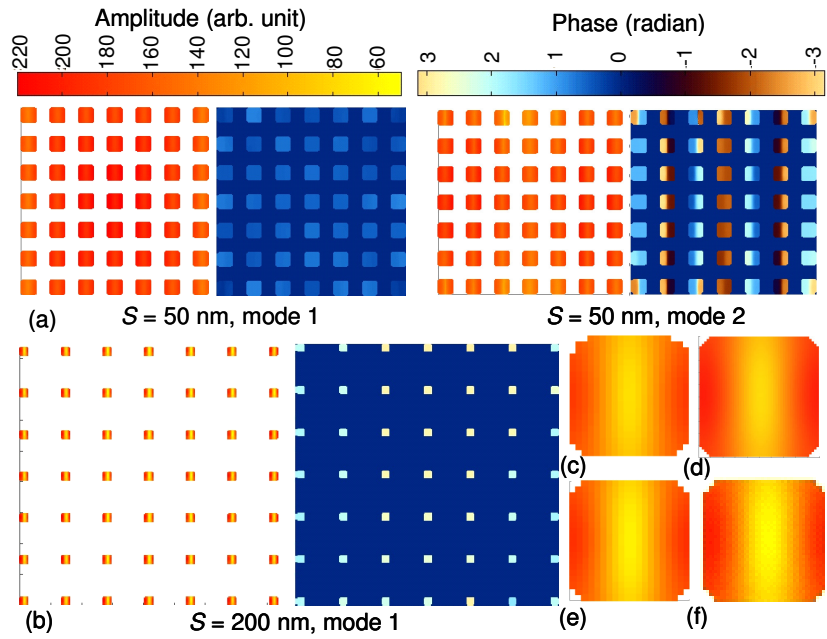

**Figure 5.** The amplitude and phase maps corresponding to different resonant frequencies are shown for the arrays with (a) $S$ = 50 nm and (b) $S$ = 200 nm. We have also simulated the amplitude maps for a single 50 nm dot with 20 nm thickness with different cell size as (c) 2.5 × 2.5 × 20 nm$^3$, (d) 1 × 1 × 20 nm$^3$, and (e) 2.5 × 2.5 × 5 nm$^3$ and compared it with (f) the central dot from the 7 × 7 array with $S$ = 200 nm. The color bars at the top of the images represent the amplitude and phase values within the images.

B. Rana et al.